\documentclass[12pt,preprint]{aastex63}
\usepackage{color}

\shorttitle{ECM emission from the Sun}
\shortauthors{White et al.}

\begin{document}

\title{Electron Cyclotron Maser Emission and the Brightest Solar Radio Bursts}

\author{Stephen M. White}
\email{stephen.white.24@us.af.mil}
\affiliation{Space Vehicles Directorate, Air Force Research Laboratory, Kirtland AFB, NM, USA}
\affiliation{Department of Physics and Astronomy, University of New Mexico, Albuquerque, NM 87106, USA}

\author{Masumi Shimojo}
\affiliation{National Astronomical Observatory of Japan, Mitaka, Tokyo 181-8588, Japan}

\author{Kazumasa Iwai}
\affiliation{Institute for Space-Earth Environmental Research, Nagoya University, Nagoya 464-8601, Japan}

\author{Timothy S. Bastian}
\affiliation{National Radio Astronomy Observatory, Charlottesville, VA, USA}

\author{Gregory D. Fleishman}
\affiliation{Center for Solar-Terrestrial Research, New Jersey Institute of Technology, Newark, New Jersey, USA}
\affiliation{Institut f\"ur Sonnenphysik (KIS), Sch\"oneckstrasse 6, D-79104 Freiburg, Germany}

\author{Dale E. Gary}
\affiliation{Center for Solar-Terrestrial Research, New Jersey Institute of Technology, Newark, New Jersey, USA}

\author{Jasmina Magdalenic}
\affiliation{Solar-Terrestrial Centre of Excellence - SIDC, Royal Observatory of Belgium, 1180 Uccle, Belgium}
\affiliation{Centre for Mathematical Plasma Astrophysics, Dept.\ of Mathematics, KU Leuven, B-3001 Leuven, Belgium}

\author{Angelos Vourlidas}
\affiliation{The Johns Hopkins University Applied Physics Laboratory, 11100 Johns Hopkins Road, Laurel, MD, USA}

\email{Draft: \today}

\begin{abstract}
This paper investigates the incidence of coherent emission in solar radio bursts, using a revised catalog of 3800 solar radio bursts observed by the Nobeyama Radio Polarimeters from 1988 to 2023. We focus on the 1.0 and 2.0 GHz data, where radio fluxes of order 10$^{10}$ Jansky have been observed. Previous work has suggested that these bursts are due to electron cyclotron maser (ECM) emission. In at least one well-studied case, the bright emission at 1 GHz consists of narrowband spikes of millisecond duration. Coherent emission at 1 GHz can be distinguished from traditional incoherent gyrosynchrotron flare emission based on the radio spectrum: gyrosynchrotron emission at 1 GHz usually has a spectrum rising with frequency, so bursts in which 1 GHz is stronger than higher-frequency measurements are unlikely to be incoherent gyrosynchrotron. Based on this criterion it is found that, for bursts exceeding 100 sfu, three-quarters of all bursts at 1 GHz and half of all 2 GHz bursts have a dominant coherent emission component, assumed to be ECM. The majority of the very bright bursts at 1 GHz are highly circularly polarized, consistent with a coherent emission mechanism, but not always 100\% polarized. The frequency range from 1 to 2 GHz is heavily utilized for terrestrial applications, and these results are relevant for understanding the extreme flux levels that may impact such applications. Further, they provide a reference for comparison with the study of ECM emission from other stars and potentially exoplanets. 
\end{abstract}


\section{Introduction}

Of the five mechanisms known to contribute to the radio emission of the Sun, electron cyclotron maser emission (ECM) is the least well studied in terms of occurrence and association with other solar phenomena. The other mechanisms (thermal bremsstrahlung, thermal gyroresonance, nonthermal gyrosynchrotron and nonthermal plasma emission) are well known and their roles in the different aspects of solar radio emission are fairly well understood \citep[e.g., see the review articles in][]{GaK04}.

Although there had been considerable interest in applications of maser forms of cyclotron emission to astrophysics previously \citep[e.g.,][and references therein]{Mel76}, modern interest was re-invigorated when \citet{WuL79} pointed out that the inclusion of semi-relativistic effects in the cyclotron resonance condition enables strong maser action for a much wider and more plausible range of electron velocity distributions than was previously thought. In particular, loss-cone velocity distributions, known to be common in planetary radiation belts, were shown to be unstable to maser action, and the ECM mechanism then became plausible as the likely source of terrestrial auroral kilometric radiation \citep[AKR; ][]{WuL79,LKW80,MRH82} and of Jovian decametric radiation \citep[e.g.,][]{HMR82}, both high-brightness-temperature radio emissions from locally strong magnetic field regions in planetary magnetospheres exhibiting a lot of structure on fine frequency and temporal scales. Subsequently \citet{ECM00} showed that a ``horseshoe'' velocity distribution was more effective for ECM and more likely to be the driver of AKR than the loss-cone distribution. 

\citet{HEK80} and \citet{MeD82} suggested that certain types of solar radio bursts might also be due to cyclotron-maser emission. \citet{Dro77} and \citet{Slo78} had reported solar radio bursts with narrow bandwidths, durations of the order of milliseconds, and inferred brightness temperatures in excess of 10$^{12}$ K, requiring a coherent emission mechanism. Furthermore, \citet{Slo78} showed that the spike bursts could be 100\% circularly polarized, which is consistent with the known properties of the cyclotron-maser emission. The other potential mechanism for high brightness temperature, highly-polarized radio bursts is plasma emission, known to be dominant at frequencies below about 300 MHz in the solar corona \citep[e.g.,][]{McL85}. Plasma emission and cyclotron emission each represent the role of fundamental physical characteristics of the solar corona: electron density in the case of plasma emission and magnetic field in the case of cyclotron emission. Plasma emission is generated at the fundamental of the electron plasma frequency $f_p$ and its (low) harmonics, where $f_p\,=\,9000\,\sqrt{n_e}$ (Hz) depends only on the ambient electron density $n_e$ (cm$^{-3}$). Plasma emission at the fundamental frequency $f_p$ can be 100\% polarized in the sense of the $o$ mode. Cyclotron maser emission occurs at the electron cyclotron frequency, $\Omega_B\,=\,2.8\times10^6\,B$ (Hz), and its harmonics, where $B$ is the magnetic field strength (in Gauss); generally, cyclotron maser emission is expected to be polarized in the sense of the $x$-mode. There has been extensive discussion of whether plasma emission or cyclotron maser emission is responsible for solar spike bursts \citep[e.g.,][]{FlM98a,Che11,CSS22}, which we will not rehash here. We note that \citet{GuZ91} found that when spike bursts occur in conjunction with Type III bursts, the spike bursts always had the opposite polarization. Since Type III bursts are known to be plasma emission and therefore likely to be $o$-mode polarized, this implies that the spikes were in the $x$ mode, which is an argument in favor of ECM. Similarly, \citet{FGN03} found that spike polarization matched that of the associated higher-frequency optically-thin gyrosynchrotron emission, indicating $x$-mode polarization. Since free-free absorption varies roughly as ${n_e}^2\,T^{-0.5}$, where $T$ is the ambient temperature, and $f_p\,\propto\,{n_e}^{0.5}$, the absorption of fundamental plasma emission increases as a high power of emission frequency, and this is believed to explain its relative rarity at GHz frequencies \citep[although it may be more likely in high-temperature environments; e.g., ][]{WhF95}. Solar spike bursts have been seen at least up to 8 GHz \citep{BSM92}.

In this paper we adopt the assumption that high brightness temperature, highly-polarized bursts at around 1 GHz are due to cyclotron maser emission. We describe a prototype example in the next section. Our focus in this paper is not on spike bursts themselves, but rather on the occurrence of solar radio bursts with these properties. One of our prime motivations is the fact that such bursts can produce perhaps the most intense natural radio emission observable from the Earth's surface \citep[of order $10^{-16}$ W m$^{-2}$ Hz$^{-1}$; e.g.,][]{Gar08,GaB21,CSS22}, and can dramatically affect the operation of global navigation satellite systems (GNSS), such as the Global Positioning System \citep[GPS; e.g., see][]{CKG08,CBG09}. A better understanding of such bursts can help to identify the physical conditions under which they arise and has implications for space weather impacts. Another important consideration is the increasing interest in cyclotron maser emission (analogous to AKR) associated with particles trapped in magnetospheres around stars such as brown dwarfs \citep{HAD06} and magnetic B stars \citep{TLL00}. Such stars exhibit pulsed radio emission, presumably due to the rotation of the stars, and the radio emission provides an important diagnostic of magnetic field properties and other features of these systems. A better understanding of the occurrence of ECM on the Sun may inform the broader astrophysical applications. To this end, in this paper we survey the solar radio burst measurements by the Nobeyama Radio Polarimeters \citep[NoRP;][and references therein]{NSS85,ShI23} at 1.0, 2.0 and 3.75 GHz since 1988. These data are available daily during Japanese daylight hours for essentially the whole period and importantly provide circular polarization measurements absent in other comparable datasets. We show from spectral considerations that ECM is, in fact, surprisingly common in solar radio bursts at 1 GHz, and that bright ECM emission lasting for an hour or more is not unusual in intense bursts at this frequency.

\section{Solar Radio Burst of 2006 December 06}

The poster child for bright L-band solar radio bursts is the 2006 December 6 X9.3\footnote{on the new NOAA ``science-quality'' scale for GOES soft X-ray flux; X6.5 on the old scale} flare, because it was well observed by several radio instruments. This event, along with several other bright bursts from the same active region, has been discussed by \citet{Gar08} and \citet{2014ApJ...789..152N}. Radio emission reached a peak flux of order 10$^6$ sfu\footnote{1 solar flux unit (sfu) = 10$^4$ Jansky = 10$^{-22}$ W m$^{-2}$ Hz$^{-1}$} at 1.4 GHz, at which level it rendered civilian GPS receivers in the western hemisphere largely unusable for about 20 minutes \citep{CKG08,CBG09}. The brightest emission in the 1.0-1.5 GHz range occurred some 45 minutes after the peak of the soft X-ray flare. Strong but rapidly fluctuating emission was seen at least across the full frequency range from 1.0 to 1.5 GHz, with orders-of-magnitude variation seen on timescales of minutes.

One of the instruments that observed the flare was the FASR Subsystem Testbed (FST) at Owens Valley Radio Observatory \citep{LGN07}, which sampled the 1.0-1.5 GHz frequency range at the (then) very high data rate of 1 Gsps\footnote{Giga (10$^9$) samples per second}. The system alternated right- (RCP) and left-circular (LCP) polarizations on a 4 second timescale. Due to limitations in the speed of writing data to disk at the time, data were acquired for 100 microseconds, then the system waited 20 milliseconds while the data were written out. Thus, timescales between 0.1 and 20 ms were not measured. Three antennas were recorded simultaneously for use as an interferometer. The recorded time-domain data were Fourier sampled to 512 frequency channels, effectively each 1 MHz wide, and accumulated over 97 $\mu$s to form each snapshot spectrum. Figure 1 shows the result of stacking these spectra in time for several consecutive 4-second intervals in right circular polarization, where the structure of the burst can be seen. Intense spikes several MHz wide and shorter than the effective time resolution of 20 ms are seen across the entire spectrum, with a higher density towards the lower end of the frequency range. Other forms of emission are present, but the flux is dominated by the spikes, which are close to 100\% RCP polarized. No temporal structure is seen in individual spikes within the fully-sampled 100-$\mu$s data periods, implying durations significantly longer than 0.1 ms \citep[consistent with the millisecond durations found in the earlier studies; e.g. ][]{Dro77,Ben86,BGI91,Rozh_etal_2008,DRK11}. No drift in frequency was seen over 100-$\mu$s data periods, but that does not mean that it would not be evident over the full duration of a spike.

\begin{figure}
\includegraphics[width=1\textwidth,angle=0,keepaspectratio]{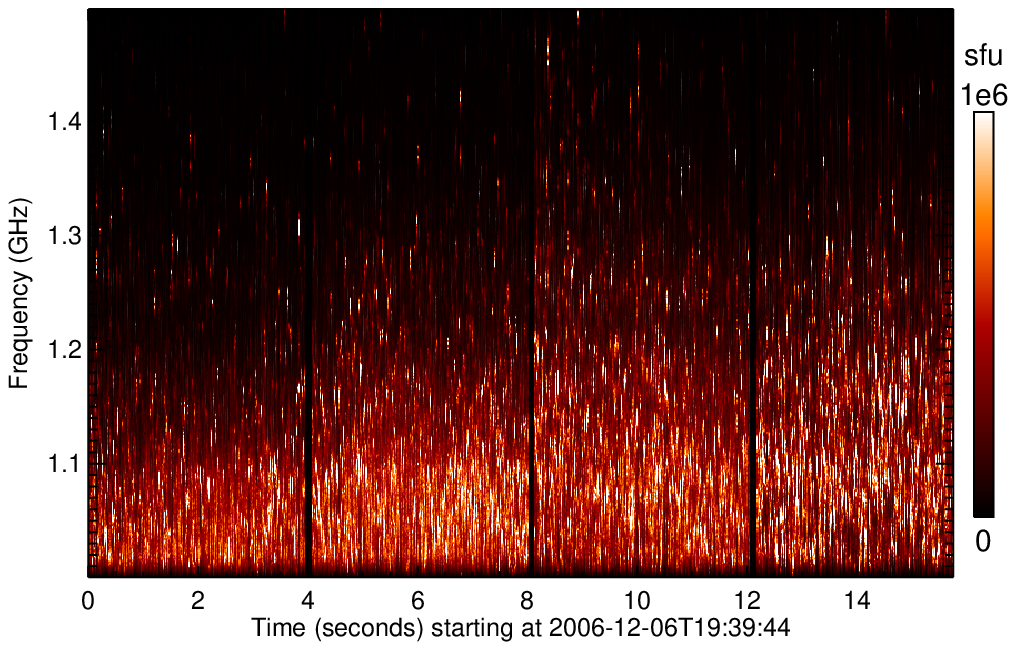}
\caption{A plot of 15 seconds of right-circular-polarization FST data sampling the correlated amplitude on a single baseline around the time of the peak at 1.4 GHz from the flare on 2006 Dec 06. Vertical black bars denote gaps in real time of 4 seconds while LCP is being recorded. The frequency resolution is 1 MHz. The data consist of 100 $\mu$s samples plotted every 20 ms. {The flux scale is shown by the color bar at right, (roughly) calibrated against the OVRO flux measurements at 1.2 GHz (see Gary 2008). The brightest spikes shown reach around 5$\times 10^6$ sfu averaged over 100 $\mu$s; the brightest 1 $\mu$s-long 1 MHz-wide samples reach around 3$\times 10^7$ sfu.} \label{figfst}}
\end{figure}

Fig. 1 demonstrates that, when measured at a coarser 1 s time resolution typical of solar radio flux monitors, individual spikes would not be fully resolved: rather, they would appear to form a structured broadband continuum. As noted by \citet{Gar08}, intense events such as this with such high densities of spikes seem qualitatively different from most flares in which spikes are observed for shorter durations at much lower densities and with much weaker contributions to the overall flux level. In particular, while spikes are seen up to 8 GHz \citep[e.g.,][]{CsB93}, extreme flux densities at the levels seen in this event seem to be restricted to frequencies below 3 GHz. Flare radio emission at higher frequencies is dominated by incoherent gyrosynchrotron emission from the nonthermal electrons that also produce flare hard X-rays. This has a spectrum that typically increases at low frequencies, where the emission is optically thick, to a peak at around 10 GHz \citep{2004ApJ...605..528N}, and decreases at (the optically-thin) frequencies above that peak \citep[e.g.,][]{Ram69,TaS70}. Thus, a straightforward method to identify events with a potentially significant ECM contribution to the radio flux at, say, 1.0 GHz, is to look for bursts in which flux does not increase with frequency as it would with a gyrosynchrotron spectrum. We explore this approach in the following sections. 

\section{Nobeyama Radio Polarimeter Catalog of Solar Radio Bursts}

NoRP (operated by the National Astronomical Observatory of Japan) provides data at 1.0, 2.0, 3.75, 9.4, 17 and 35 GHz, at 1 s time resolution\footnote{NoRP data at 0.1 s time resolution is also available for bright bursts.} in both left and right circular polarization (see \url{http://solar.nro.nao.ac.jp/norp/index.html}). As noted above, strong circular polarization is often a characteristic of ECM emission, and we are interested in determining how often high degrees of polarization are seen in bursts, so NoRP provides the best data source for such a study. In addition, the NoRP data do not suffer from artificial limits on flux reporting and occasional saturation that has affected Radio Solar Telescope Network (RSTN) data at times \citep[see discussion of this issue for RSTN data in][]{GKL17}. A catalog of NoRP solar radio bursts from 1988 is provided at \url{http://solar.nro.nao.ac.jp/norp/html/event/} \citep[used by][for their study of NoRP burst flux distributions]{SHT12}, but it ceased to be maintained after early 2015 due to resource issues. Further, inclusion of radio bursts did not occur in an organized fashion in the early years, so there are missing bursts that would likely have been recorded in more recent years. Since we wish to have as complete a survey of ECM bursts as possible, we have therefore generated a new catalog of NoRP bursts from 1988 to 2023 using a combination of automated detection and manual\footnote{By SW} verification \citep[also used for a statistical study of ``cold'' solar flares;][]{2023ApJ...954..122L}. The new catalog should be reasonably complete for bursts with flux exceeding 50 sfu in the 1-17 GHz range; it contains 3818 events, compared to 856 events in the previous catalog. Note that this is a radio-selected catalog of bursts, not flare-selected, so flares in which multiple bursts occurred well separated in time, particularly at 1.0 GHz, may be represented more than once (e.g., the flare of 2006-12-13 discussed further below). \citet{ShI23} describe the history of the NoRP data and issues at each frequency over the years that can result in spurious features, and we have tried to take these into account in the analysis. Generally a clear burst detection at more than one frequency was required for the burst to be included in the list. However, bursts are included at 1.0 and 2.0 GHz in the absence of higher-frequency emission if they are associated with a clear prior event: the results of this paper will justify that approach. 

\begin{figure}[t]
\centering
\includegraphics[width=1.00\textwidth,angle=0,keepaspectratio]{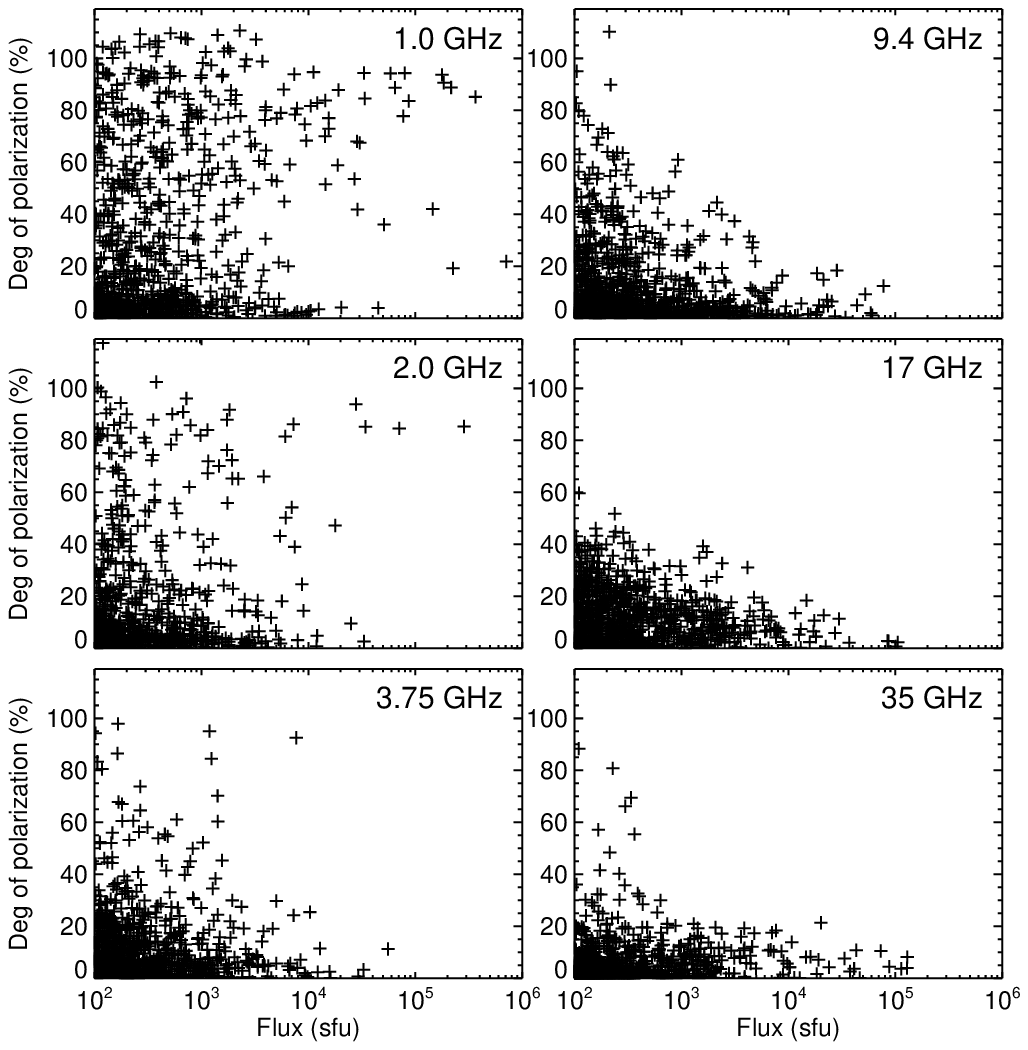}
\caption{A plot of degree of circular polarization versus peak flux for NoRP events in excess of 100 sfu. The degree of polarization is calculated {at the time of the peak Stokes I flux}. The numbers of events plotted are 833, 733, 823, 1160, 876 and 1276 at 1.0, 2.0, 3.75, 9.4, 17 and 35 GHz, respectively. At 35 GHz the noise level in the data can be quite high, particularly in poor weather conditions, so we expect that many of the weaker detections there are not reliable. \label{figpol}}
\end{figure}

\section{Survey of ECM in Solar Radio Bursts}

We focus on using the burst emission at the 3 lowest NoRP frequencies since the brightest ECM bursts so far have occurred at frequencies below 2 GHz. For an initial survey, in Figure 2 we plot the degree of polarization versus peak flux at each frequency for all bursts where the flux at that frequency exceeds 100 sfu. Here, the degree of polarization is measured {at the time of peak Stokes I flux}. Several things are clear from this figure. Of the six frequencies, 1.0 GHz shows the largest number of events with both large flux and a high degree of polarization. 2.0 GHz also exhibits many events with large polarization but generally lower fluxes. The higher frequencies generally have low polarization in all high-flux events: this is not surprising for gyrosynchotron emission, since high fluxes will often be from an optically thick source, thus having a lower polarization. Calculations also show that the degree of polarization of optically-thin gyrosynchrotron emission decreases as the frequency increases \citep[e.g., ][]{DuM82}, although it can still be rather large in the case of anisotropic angular distributions of nonthermal electrons \citep[see, e.g., figures\,7--9 in][]{Fl_Meln_2003b}. High degrees of polarization at low flux levels may indicate high noise levels in V that produce unrealistic values.

As mentioned above, both coherent and incoherent emission can contribute at 1.0 and 2.0 GHz, but incoherent gyrosynchrotron emission should in most cases have a clear signature of flux increasing with frequency up to a single spectral peak. The Tsytovich-Razin effect, which suppresses gyrosynchrotron emission at frequencies not far above the plasma frequency, enhances this effect in the low-frequency region of the spectrum \citep{Ram69}. An exception to this picture is for events occurring beyond the visible limb of the Sun. The location of the gyrosynchrotron spectral peak is strongly correlated with the magnetic field strength in the source \citep{DuM82}, and field strength generally decreases with height above the solar surface \citep[see, e.g., Figure\,1 in][]{2023ApJ...943..160F} such that higher frequency emission from strong fields low down will not be visible in occulted events, shifting the spectral peak to lower frequencies. However, the number of events in which this is likely to be an issue is small, and none of the strong events that we highlight below are in this category. Thus, we identify bursts in which the 1.0 GHz flux is larger than the 3.75 GHz flux as candidate ECM events.

\begin{figure}[t]
\centering
\includegraphics[width=0.95\textwidth,angle=0,keepaspectratio]{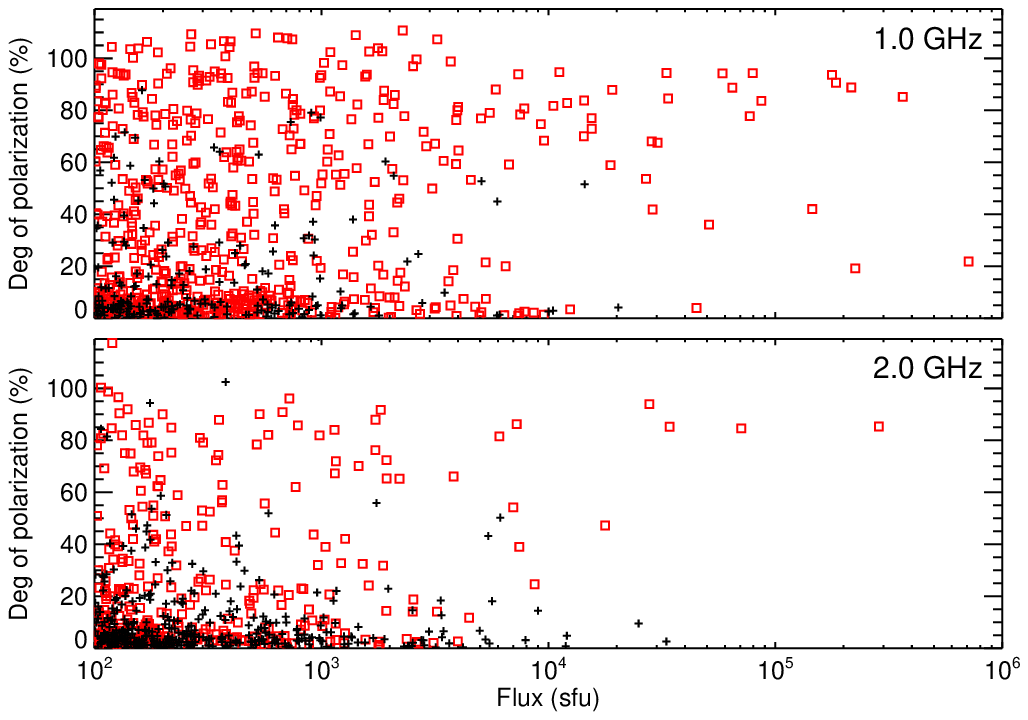}
\caption{A plot of degree of circular polarization {(at the peak of the I flux)} versus peak flux for NoRP events in excess of 100 sfu at 1.0 and 2.0 GHz. Events are labelled according to whether the flux at the frequency plotted is greater than (red open squares) or less than (black plus symbols) the corresponding peak flux at 3.75 GHz. A standard solar radio burst gyrosynchrotron spectrum would generally have flux increasing from 1.0 to 2.0 to 3.75 GHz, and thus be among the black symbols.\label{fig12}}
\end{figure}

Figure \ref{fig12} repeats the plots of Figure \ref{figpol} for 1.0 and 2.0 GHz, but now separates events in which the spectrum increases with frequency from 1.0 to 2.0 to 3.75 GHz (black plus symbols) from those in which the spectrum decreases with frequency (red open squares). It is striking that most of the high-polarization, high-flux events at 1.0 GHz have decreasing spectra, and thus are unlikely to be gyrosynchrotron emission. At 1.0 GHz, 628 events above 100 sfu have a larger flux at 1.0 GHz than at 3.75 GHz, with 205 having lower flux at 1.0 GHz. Thus, fully three-quarters of the $>$100 sfu events at 1.0 GHz can be ECM emission. Visual inspection of the NoRP lightcurves for these events confirms that in most cases the 1.0 GHz lightcurve is clearly not compatible with the higher frequency lightcurves that can be attributed to gyrosynchrotron emission. Three-quarters is a significantly larger fraction than the direct degree of association found by \citet{BGC05}: in a dedicated survey of coherent emission in a sample of 201 flares detected in hard X-rays by the RHESSI satellite, they found decimeter spikes in only 21 events, or 10\% of the sample. In most of those cases the decimetric spikes coincided in time with the peak in hard X-rays, leading to the inference that they are the coherent emission most closely associated with the main flare energy release \citep[e.g., see discussion in ][]{BaB09}. However, since we have selected events with 1.0 GHz fluxes greater than 100 sfu, this likely biases the fraction found to have coherent emission. At 2.0 GHz, there are 353 NoRP events over 100 sfu with flux greater than at 3.75 GHz, and 380 events with smaller flux: thus the incidence of dominant coherent emission at 2.0 GHz is significantly smaller than at 1.0 GHz, as inferred previously. Some of the anomalous events in which there is high polarization at 2.0 GHz but less flux than at 3.75 GHz are found to be events in which the 2.0 GHz peak is much larger than the coeval 3.75 GHz flux, but then 3.75 GHz peaks at a very different time; others may be due to RFI at either 3.75 GHz or 2.0 GHz \citep{ShI23}. 

\begin{deluxetable*}{lcrcrrrrr}
\tablecolumns{9}
\tablewidth{0pc}
\tablecaption{\label{tab_bright} NoRP solar radio bursts exceeding 10$^4$ sfu at 1.0 GHz}
\tablehead{\colhead{Date} & \colhead{SXR peak} & \colhead{SXR\tablenotemark{a}} & \colhead{1 GHz} & \colhead{1.0 GHz} & \colhead{2.0 GHz} & \colhead{3.75 GHz} & \colhead{1.415 GHz\tablenotemark{b}} & \colhead{0.61 GHz\tablenotemark{b}} \\ 
\colhead{}  & \colhead{time} & \colhead{flare} & \colhead{peak time} & \colhead{peak (sfu)} & \colhead{peak (sfu)} & \colhead{peak (sfu)} & \colhead{peak (sfu)} & \colhead{peak (sfu)}
}
\startdata
1989-06-04 & 02:21:00 &  M1.4 & 02:15:02 &  10515 &   3594 &    106 &   5100 & 820 \\
1989-08-15 & 03:17:00 &  X1.4 & 03:38:01 &  28803 &  11926 &  16096 &  77000 & 28000 \\
1989-11-26 & 23:53:00 &  X1.6 & 00:15:01 &  15531 &    673 &    390 &   3000 & 49000 \\
1990-04-15 & 02:59:00 &  X2.1 & 04:40:26 & 711276 &   4312 &   1456 &  78000 & 303000 \\
1990-07-30 & 07:35:00 &  M6.4 & 07:59:35 &  18796 &   1933 &   1353 &   2600 & 2400 \\
1991-03-23 & 04:28:00 &  M9.9 & 02:41:44 &  14331 &     22 &     20 &    800 & 9700 \\
1991-05-16 & 07:02:00 &  X1.3 & 06:58:49 &  15498 &  33140 &    712 &     94 & 9700 \\
1991-06-06 & 01:07:00 & X30.2 & 01:06:18 &  20332 &  33106 &  55690 &  24000 & 13000 \\
1991-06-09 & 01:43:00 & X15.1 & 02:39:09 &  77220 &   7252 &   2485 &  57000 & 21000 \\
1991-06-15 & 08:17:00 & X34.7 & 08:18:14 &  14447 &  12064 &  15771 &  46000 & 5000 \\
1991-08-25 & 01:13:00 &  X3.0 & 01:32:25 &  12470 &   1931 &    106 &  26000 & 680 \\
2001-04-09 & 02:56:00 &  C9.1 & 02:47:28 &  86902 &    965 &    153 &  27000 & 82000 \\
2001-10-19 & 01:05:00 &  X2.3 & 01:24:34 &  28538 &    176 &   1189 &   8300 & 32000 \\
2001-11-22 & 23:30:00 &  X1.4 & 22:47:02 &  12086 &    841 &   1181 &  53000 & 3500 \\
2002-03-18 & 02:31:00 &  M1.4 & 03:03:19 &  14373 &    129 &     54 &  27000 & 2900 \\
2002-04-21 & 01:51:00 &  X2.1 & 02:03:54 & 145440 &   8692 &   1022 & $>$100000 & 13000 \\
2003-05-28 & 00:27:00 &  X5.1 & 00:27:23 &  30279 &   1869 &   1716 &   2500 & 81000 \\
2006-12-13 & 02:40:00 &  X4.9 & 02:28:08 & 364894 &  34209 &   3014 & $>$150000 & 120000 \\
2011-02-15 & 01:56:00 &  X3.1 & 02:34:17 & 185170 &   1457 &    595 &  59000 & 47000 \\
2012-03-05 & 04:09:00 &  X1.6 & 04:28:38 & 216365 &  17794 &   4651 &  25000 & 650000 \\
2012-03-07 & 00:24:00 &  X7.7 & 01:07:27 &  33732 &   7444 &   7278 &   5300 & 430000 \\
2022-01-29 & 23:32:00 &  M1.1 & 23:30:58 &  19113 &    787 &        &   9000 & \\
2022-06-13 & 04:07:00 &  M3.4 & 03:53:11 &  64690 &  27857 &     31 &  98000 & 3500 \\
\enddata
\tablenotetext{a}{Flare class is defined on the modern ``science-quality'' soft X-ray scale. 
In 1991 and earlier, reported GOES fluxes saturated at X15.7: the 3 flares from
June 1991 were saturated, and the values shown here are from the analysis
by \citet{HCW24}}
\tablenotetext{b}{From Radio Solar Telescope Network (RSTN, operated by the US Air Force) data. 
Where 2 observatories detected the same burst, the larger reported flux was used.}

\end{deluxetable*}

\section{The Brightest 1.0 GHz bursts}

Table \ref{tab_bright} lists the 23 flares in the NoRP catalog where the 1.0 GHz flux exceeds 10$^4$ sfu\footnote{A number of these events were also discussed by \citet{CWB11}, who also list a sample of bright events in the 1.0-1.6 GHz range.}. Here we have in effect consolidated multiple radio bursts from the NoRP catalog that we believe to be associated with the same flare. A flare from 1991-03-22 has been excluded, since the excess above 10$^4$ sfu is limited to a single 1-second integration and is likely to be spurious. We note that for a gaussian source of full-width at half-maximum of 20\arcsec, this flux threshold would require a brightness temperature of 1.6$\times 10^{11}$ K, so this flux level cannot plausibly be produced by any incoherent emission mechanism, thermal or nonthermal. Plots of the temporal behavior of the NoRP fluxes, the degree of polarization at 1.0, 2.0 and 3.75 GHz, and the corresponding SXR profile are presented in Figure \ref{fig_flares} in Appendix A. We omit the flare of 1991 June 6 where the radio flux rises from 1.0 to 3.75 GHz, so there is no evidence for ECM emission in that event.

Many of the flares in Table \ref{tab_bright} are large X-class flares with long durations in soft X-rays, but there are also several M1- or C-class flares (i.e., C flares on the legacy flux scale employed by NOAA at the time of the events), indicating that conditions for production of bright 1.0 GHz emission favor, but do not require, a large flare. The fluxes of the 3 lower frequencies are plotted on a logarithmic scale in the top panels of Fig. \ref{fig_flares}, and this often obscures just how much brighter the 1.0 GHz emission is compared to 3.75 and/or 2.0 GHz. The fluxes reported in Table \ref{tab_bright} demonstrate this: in 16 of the 25 bursts, the 1.0 GHz flux is at least an order of magnitude larger than the 2.0 GHz flux. However, there are a number of bursts where 2.0 GHz is also very bright, implying coherent emission at that frequency also. For reference, we also list the 0.6 and 1.4 GHz fluxes from observations by RSTN. Since the list of bursts was selected based on 1.0 GHz flux, it is not surprising that the 1.4 and 0.6 GHz fluxes are usually smaller than the 1.0 GHz flux, but this is not always the case, and extreme fluxes are often observed at each of these frequencies as well. The log scaling also tends to obscure the degree of variability typical of these bursts: order of magnitude variation on timescales of minutes appears to be common.

Of critical interest for understanding this class of bursts is when the bright emission occurs in relation to the impulsive phase, and how long it lasts. As noted above, in the event of 2006 Dec 06 the brightest emission at 1.4 GHz occurred about an hour after the impulsive phase. Fig. \ref{fig_flares} shows a range of behavior, but generally the bright emission is long lasting (several hours in the examples of 1990-04-15, 2002-04-21, 2006-12-13, 2011-02-15), and can peak over an hour after the SXR peak (e.g., bursts of 1990-04-15, 1991-03-23, 1991-06-09, 1991-08-25). It is notable that many of the X-ray flares accompanying the bright 1 GHz emission are long-duration events. We note that \citet{KGW04} carried out a detailed study of the 2002-04-21 event: they could not identify EUV signatures corresponding to the bright 1 GHz emission (which they then attributed to plasma emission), but they did find that nonthermal 17 GHz emission and possibly nonthermal hard X-rays continued during the period of ECM emission.

\begin{figure}[t]
\plotone{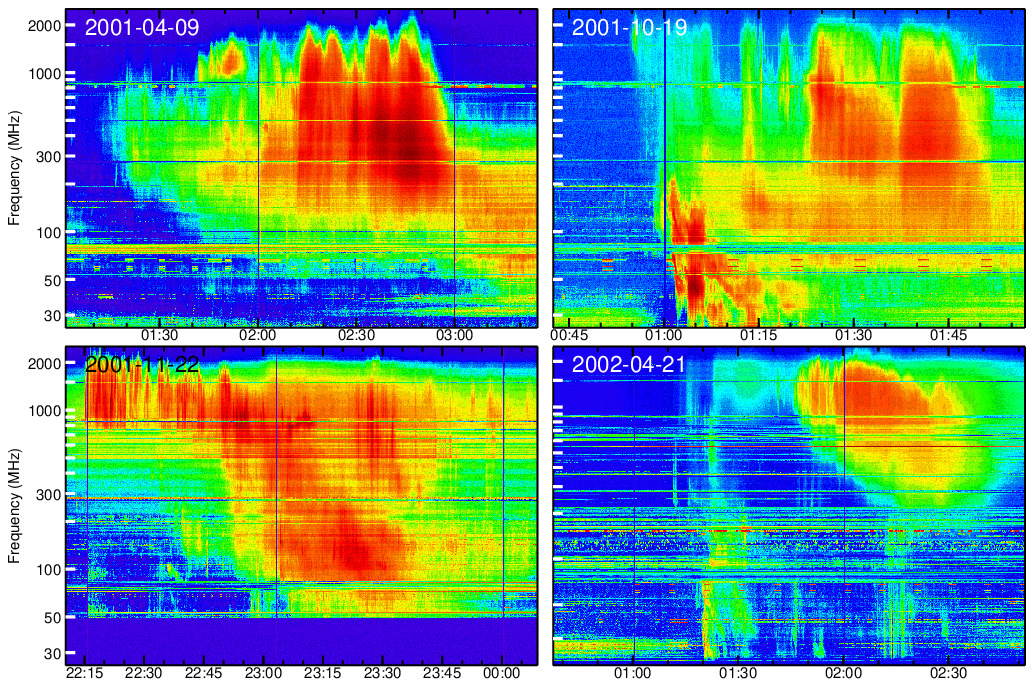}
\plotone{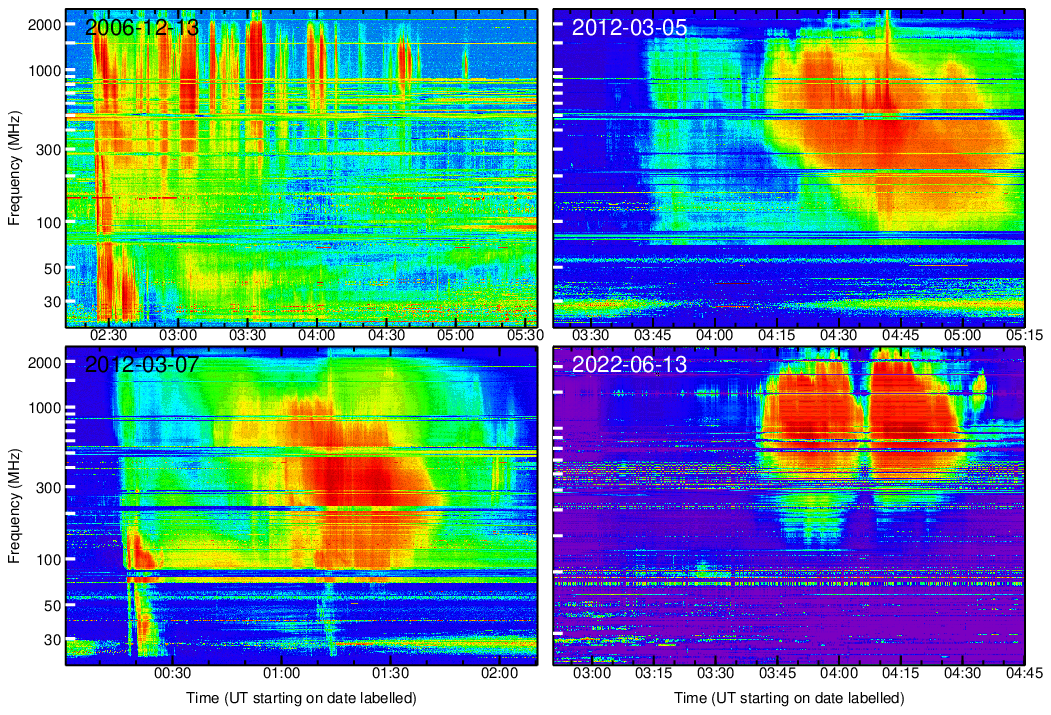}
\caption{Dynamic spectra of selected bright bursts from the Hiraiso and Yamagawa radio spectrographs operated by NiCT (see text). The plotted time ranges generally match the corresponding plots in Fig. \ref{fig_flares}. The data files are provided in units of dB. In the color table used, blue is the weakest emission, green is intermediate and red is the brightest.\label{fig-hiras}}
\end{figure}

The frequency extent and time variability of the bright emission can be seen for 8 selected events in Figure \ref{fig-hiras}. This shows dynamic spectra from the Hiraiso \citep[1996-2016; see][]{KIW95} and Yamagawa \citep[2016-present; see ][]{IKI17} radio spectrographs operated by the National Institute of Information and Communications Technology (NiCT) of Japan. These spectra extend from below 100 MHz to at least 2.0 GHz, and thus cover the frequency range of interest here. Since the data are provided in units of dB, we do not use these plots for a quantitative discussion, but they do show the extent (in frequency) of bright emission. Consistent with the high 0.6 and 1.4 GHz fluxes in Table \ref{tab_bright}, these events typically have bright emission over a wide range of frequencies lasting on the order of an hour or more. The bright emission often extends up to 2.0 GHz, but the flux values in Table \ref{tab_bright} indicate that it generally does not extend to 3.75 GHz; however, at least for 2006-12-13, \citet{WYL08} and \citet{TWW21} report that spikes were seen at that frequency. 

Since frequency maps to magnetic field strength for ECM emission, the broadband nature of the bright features in Fig. \ref{fig-hiras} implies that in these very bright bursts the conditions for bright ECM emission typically occur simultaneously over a range of magnetic field strengths, up to a factor of 10.  If ECM emission is at the fundamental frequency $\Omega_B$ then 2 GHz requires 700 G magnetic fields, while 300 MHz requires 100 G fields. This wide range of field strengths seems to imply that a large volume of the corona is subject to maser action simultaneously. For ECM emission the condition $f_p\,<\,\Omega_B$ must then be satisfied throughout the entire volume, which seems to argue against an association with the dense post-flare loops generally seen in the decay phase of long-duration flares. \citet{CWB11} argued that there could still be locations within the loop system where electric fields are present and simultaneously accelerate electrons while evacuating ducts, as they are believed to do in AKR source regions. \citet{MeW16} also investigated this scenario and concluded that horseshoe-driven ECM was possible, albeit for spike bursts occurring in conjunction with hard X-rays. Note that \citet{WMD83} show that maser-favoring velocity distributions resembling a ``horseshoe'' would arise naturally in solar coronal loops following a sudden localized injection of accelerated particles. The electric fields that accelerate particles in AKR source regions are confined to vertically very narrow ``double'' layers, of order 10 Debye lengths across, with kilovolt potential drops: in solar post-flare loops the Debye length is typically less than 1 mm, tiny compared to the typical length scales of the loops, so it seems unlikely that such small structures could accelerate enough electrons to explain the bright solar bursts. Thus a direct analogy with the generation of AKR is probably not appropriate for the large bursts discussed here. Since relatively few post-flare loop systems produce such bright ECM emission, additional conditions must be required \citep[a conclusion also reached by][]{CSS22}. The event of 2006-12-13 has repeated pulses of very bright emission occurring over several hours, suggesting that conditions for bright emission can somehow form, decay and re-form well after the main energy release in the impulsive phase of the flare is long over.

A significant feature of Fig. \ref{fig-hiras} is the morphological similarity of the bright features in the dynamic spectra of all these events. We do not have millisecond-resolution observations of spikes in all these events, but we know that 2006 Dec 6 and 13 were dominated by spikes: the similarity of the spectral features argues that the other bright bursts were likely also dominated by spikes.

Another important property is polarization. The degree of polarization at the three lower frequencies is plotted in the middle panels for each event in Fig. \ref{fig_flares}. Consistent with the results shown in Fig. \ref{figpol}, the polarization is typically high, but rarely at 100\%. The highest degrees of polarization seem to occur when the flux is highest (e.g., 1990-07-30, 1991-06-909, 2001-04-09, 2001-10-19, 2006-12-13, 2012-03-07). \citet{GuZ91} studied the polarization of solar radio spike bursts and found a wide spread in the degree of polarization, with an apparent center-to-limb variation such that polarization was highest near disk center. Although the two bursts in our sample that have consistently lower polarization (1991-06-15, 1991-08-25) are both limb events, there are other limb events (e.g., 2002-04-21) with high degrees of polarization, and within a given event the degree of polarization can vary greatly. If ECM is intrinsically 100\% polarized, this would imply that either the polarization has been affected by propagation effects, or else a significant component of the coherent emission in the NoRP data is not due to ECM. 

An interesting feature of Fig.\,\ref{fig-hiras} is the absence of bright low-frequency emission in the range traditionally associated with plasma emission (below 300 MHz) at the times when the 1.0 GHz emission is brightest. Low-frequency radio bursts can be seen in the impulsive phase in several of the events in Fig. \ref{fig-hiras} (notably 2001-10-19, 2002-04-21, 2006-12-13 and 2012-03-07), but there is no such emission later when the 1.0 GHz emission is bright. The absence of emission such as Type III bursts from electron beams on open field lines, and the clear lower-frequency bound to the frequency range of bright emission in most events, suggests that the conditions for the bright ECM emission are confined to regions of closed magnetic field.

\begin{figure}[t]
\centering
\includegraphics[width=0.90\textwidth,angle=0,keepaspectratio]{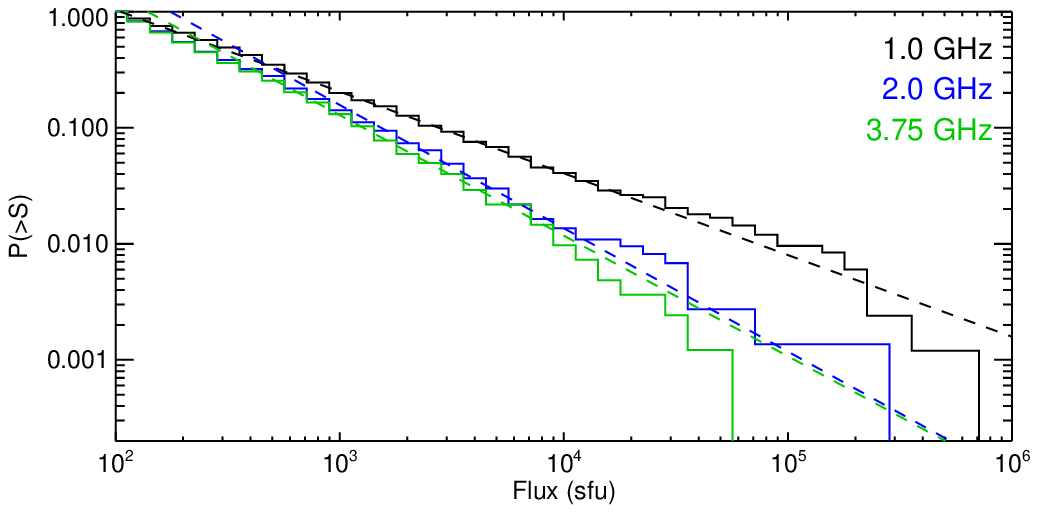}
\caption{Cumulative probability distributions for the NoRP peak flux at 1.0 (black), 2.0 (blue) and 3.75 (green) GHz, for bursts larger than 100 sfu. The corresponding dashed lines are power-law fits to the 1000-10000 sfu range.
\label{figcumul}}
\end{figure}

\section{Extreme-flux limits}

An issue of some importance is the question of just how bright ECM bursts can be in this frequency range. This is relevant for benchmarking the impacts of space weather \citep[e.g.,][]{SWB18,SWB19}. \citet{Ril12} and \citet{RiL17} describe an approach for estimating the probability of event occurrence exceeding a specific threshold given a power-law distribution of fluxes.

Those papers note the value of using cumulative distributions for analysis, i.e., distributions representing the number of events observed above a given flux value. Figure \ref{figcumul} plots the cumulative probability distributions of fluxes above 100 sfu from the NoRP burst catalog at 1.0, 2.0 and 3.75 GHZ, together with power-law fits derived from the range 10$^3$ - 10$^4$ sfu. A single power-law (index -0.70, fitted from 132 events lying between 10$^3$ and 10$^4$ sfu) fits the 1.0 GHz cumulative distribution across most of the range from 100 to 10$^5$ sfu, with an excess above the fit at around 10$^5$ sfu. As noted above, an individual flare may be represented by multiple entries in the NoRP catalog, particularly at 1.0 GHz, and this factor may contribute to the excess at $\sim\,10^5$ sfu. The fits at 2.0 and 3.75 GHz are similar to each other and somewhat steeper than the 1.0 GHz distribution (power-law indices -1.06 and -1.04 from 94 and 100 events, respectively), but they are above the actual distributions at lower flux levels, indicating that a single power law is not appropriate across the entire range of fluxes. It is possible that the 1.0 GHz fluxes are dominated by ECM events down to 100 sfu, whereas the lower flux regime at 2.0 and 3.75 GHz has a larger fraction of gyrosynchrotron events contributing with a different distribution. The 2.0 GHz distribution stays close to the power-law fit out to 2$\,\times\,10^5$ sfu, while the 3.75 GHz distribution drops below the power law at fluxes beyond $10^4$ sfu. The formal uncertainties in the power-law index fits are of order 10\%. The power-law indices for the cumulative distributions correspond to power-laws of -1.70, -2.06 and -2.04 for the flux occurrence distribution functions above 100 sfu. \citet{NGL02} fitted power-laws to 40 years of radio bursts above 100 sfu at various frequencies, and found occurrence distribution function indices of -1.83 for both 1.0-1.7 GHz and 2.0-3.8 GHz ranges. \citet{GKL17} analyzed the distributions of bursts in the RSTN data (which, they note, suffer from reporting issues that produce under-sampling), and, fitting over the full range of fluxes above 100 sfu, they found power-law indices of -1.88, -1.82 and -2.01 at 0.6, 1.4 and 2.7 GHz, respectively. The differences in power-law fits between the different surveys can be attributed at least in part to the different flux ranges being used for the fits.

If we assume that the fitted power laws continue to apply beyond the fitted flux range, we can apply the formalism of \citet{Ril12} to estimate the incidence of high fluxes. We assume an average coverage of 10 hours per day for the NoRP observations from 1988 to 2023 (36 years) that were used to derive the radio burst catalog: scaling this to 100 years of 24-hour coverage yields a factor of 6.67. We find that we would expect to see 9 bursts over 10$^6$ sfu at 1.0 GHz in 100 years, 1.4 at 2.0 GHz, and 0.5 at 3.75 GHz. At the probability level of 50\%, the largest bursts that might be seen in 100 years are $5\,\times\,10^7$ sfu at 1.0 GHz, $3\,\times\,10^6$ sfu at 2.0 GHz, and $1\,\times\,10^6$ sfu at 3.75 GHz. 

However, if the distributions do not follow the power-law fits beyond 10$^6$ sfu, then these numbers will be overestimates. \citet{HCW24} provide an extensive discussion of fitting the roll-over of the flare soft X-ray flux distribution at high flux levels, but that roll-over is much more pronounced than those in Fig. \ref{figcumul} and we do not attempt such a fit for the radio flux distributions. The nature of Fig. \ref{figfst} suggests that individual maser spikes saturate after a few milliseconds but that conditions for maser action can re-establish somewhere else in the emitting volume quickly. There may be limitations on this process that could affect the upper limit to fluxes, e.g., if formation of a maser in a certain region limits the simultaneous or subsequent formation of maser action within a volume around that location, possibly because of the effect of the maser radiation on the velocity distribution in the volume that it irradiates. Better theoretical understanding of the mechanism of these bursts is required to further address the question of practical upper limits to extreme fluxes. Note that \citet{GKL17} carried out an extreme-event analysis based on their fits to the RSTN burst data, and estimated an upper limit to the extreme flux at 1.4 GHz of 3.2 $\times\,10^6$ sfu over a 100-year period.

\section{Summary}

In this paper we investigate the incidence of electron cyclotron maser emission in solar radio bursts. We use a revised catalog of radio bursts observed by the Nobeyama Radio Polarimeters from 1988 to 2023, consisting of 3800+ radio bursts, and focus primarily on the 1.0 GHz data since previous work has suggested that ECM emission can be prominent there. We can distinguish coherent emission at 1 GHz from incoherent gyrosynchrotron emission based on the radio spectrum: gyrosynchrotron emission at 1 GHz usually has a spectrum rising with frequency, so bursts in which 1 GHz is stronger than higher-frequency measurements are unlikely to be incoherent gyrosynchrotron. Based on this criterion, three-quarters of all bursts exceeding 100 sfu at 1 GHz, and half of the bursts at 2 GHz, have a significant coherent emission component, assumed to be ECM. The advantage of NoRP data is that the degree of circular polarization in the radio burst is also measured: the catalog shows that most very bright bursts at 1 GHz are highly polarized, consistent with a coherent emission mechanism. The brightest emission seems to be most highly polarized, but it does not appear that the coherent emission is always 100\% polarized. 

As suggested by the study of the famous 2006-12-06 burst, the brightest 1 GHz emission tends to occur well after the impulsive phase of the flare, and can last for an hour or more. This implies that the source of the energetic electrons that radiate the brightest 1 GHz bursts is not the main energy release in the flare: these radio bursts are apparently fundamentally different from the better-studied spikes in the impulsive phase of flares, which are often associated with the electrons that produce flare hard X-ray emission \citep[e.g.,][]{GAB91,BGC05,BaB09} and/or microwave emission \citep{FGN03}. A pronounced characteristic of the very bright bursts is that they extend across a surprisingly wide frequency range for quite a long period: dynamic spectra show the structure associated with ECM often extending down at least to 300 MHz and up to 2.0 GHz, but not often beyond 2.0 GHz. The corresponding range of coronal magnetic field strengths (assuming radio emission at $f\,=\,\Omega_B$) is 100 to 700 G: why ECM should favor this range of field strengths is a question needing further study. Note that the delay in the bright ECM emission means that there is little connection to the gyrosynchrotron emission produced by a flare, thus ruling out application of the model invoked by \citet{FlM98a} to explain spikes during the impulsive phase. 

Six events in the NoRP catalog have 1.0 GHz fluxes exceeding 10$^5$ sfu. The distribution of peak fluxes at 1.0 GHz over the period of the NoRP burst catalog obeys a power law out to about 10$^6$ sfu. This can be extrapolated to an estimate of the largest solar radio burst that can be expected: in a 100-year period, at 1.0 GHz this could be as high as 5$\,\times\,10^7$ sfu. A flux at this level extending from 0.3 to 2 GHz would have severe impacts on cell-phone reception and global navigation systems across most of the sunlit hemisphere of the Earth. However, this estimate relies on the power law continuing to larger fluxes: if there is a saturation effect that truncates the flux distribution, then the upper limit will not be this high. Similarly, peak fluxes as high as 3$\,\times\,10^6$ and 1$\,\times \,10^6$ sfu could be expected at 2.0 and 3.75 GHz, respectively.

It is worth noting that a 10$^6$ sfu radio burst on a star 10 pc distant would produce 2.3 mJy of radio flux at Earth, which is easily detectable by modern radio telescopes. Thus the brighter solar ECM bursts would be readily detectable if occurring on nearby stars, and indeed there have been high-sensitivity observations suggesting analogs of spike bursts on dMe stars \citep[e.g.,][]{ALA97,ZTZ23}. By analogy with terrestrial AKR and Jovian DAM, ECM is also the favored explanation for the coherent rotationally-modulated radio bursts observed in brown dwarf and magnetic B stars which are believed to have magnetic field structures dominated by a dipole component, much like planetary magnetospheres. Then an important question is whether ECM from solar flares is relevant for such systems where the ECM is not necessarily associated with flares. AKR exhibits a lot of fine structure, usually with pronounced frequency drift \citep[e.g.,][and references therein]{BMH88,YeP22,TFP23} that is attributed to motion of the source region up or down magnetic field lines with particularly low plasma density (hence $f_p\,<\,\Omega_B$) in the auroral regions of the Earth's magnetosphere. Electric fields play a major role in both accelerating electrons into maser-favoring ``horseshoe'' velocity distributions and evacuating the field lines \citep{ECM00}, and in turn the auroral electric fields are affected by the impact of the solar wind on the magnetosphere.  Jovian DAM displays a wide range of fine structure, some associated with the motion across the magnetosphere of the flux tube attached to the Io satellite, and others associated with solar wind parameters \citep[e.g.,][]{RZR97,CHS14,PRR18}. Dynamic spectra of those emissions do not generally look like Fig. \ref{figfst}. Solar spike bursts can occur in chains that drift in frequency, but individual decimetric spikes are narrowband \citep[bandwidth of order a few percent at most, e.g.,][]{GuB90,CsB93,Rozh_etal_2008,NFG08,DRK11,2014ApJ...789..152N}, which is consistent with theoretical consideration of ECM \citep{Fl_2004_AL,Fl_2004,Rozh_etal_2008}.  But in solar bursts any drift in frequency is very limited, unlike AKR where drifts of 10\% or more in frequency are common in discrete features. Thus the relationship between the brightest solar ECM bursts and planetary/brown dwarf radio emission is not clear, and requires further investigation.

\acknowledgments
SW thanks AFOSR for support for basic research through LRIR 23RVCOR003. The views expressed herein are those of the authors and do not reflect official guidance of the US Government or Dept. of Defense. The appearance of external hyperlinks does not constitute endorsement by the US Dept. of Defense of the contents of such links. DG and GF acknowledge support from NSF grants AGS-2130832 and AGS-2121632 and NASA grant 80NSSC23K0090 to New Jersey Institute of Technology.

\facilities{This work uses data from the Nobeyama Radio Polarimeters (NoRP) operated by the National Astronomy Observatory of Japan (NAOJ).}

{\large\it Data availability:} NoRP data are available publicly from the NAOJ website at \url{http://solar.nro.nao.ac.jp/norp/index.html}. The catalog of NoRP solar radio bursts used for this study is available on Zenodo at \dataset[\doi{10.5281/zenodo.10719814}]{\doi{10.5281/zenodo.10719814}}. Data from the Hiraiso and Yamagawa radio spectrographs are publicly available at \url{https://solobs.nict.go.jp/radio/cgi-bin/MainDisplay.pl}.

\clearpage


\clearpage

\appendix

\section{Event Plots}

The following figures show the temporal behavior of the radio emission in the 1.0 GHz-bright events listed in Table \ref{tab_bright}. In each case the upper panel shows the 1.0, 2.0 and 3.75 GHz data on a logarithmic plot; the middle panel shows the corresponding (absolute) degree of circular polarization of those three frequencies; and the bottom plot shows the 3.75, 9.4, 17 and 35 GHz data on a linear scale, for comparison, along with the time profile of the GOES soft-X-ray emission from the associated flare. Note that the radio emission in the bottom panel is generally believed to be gyrosynchrotron emission from nonthermal electrons accelerated in the flare in the impulsive phase, with possible contributions from thermal bremsstrahlung emission in the decay phase. The emission at 1 GHz, and occasionally that at 2.0 and 3.75, requires brightness temperatures in excess of 10$^{11}$ K for any plausible source size, which requires a coherent emission.

\begin{figure}[b]
\plotone{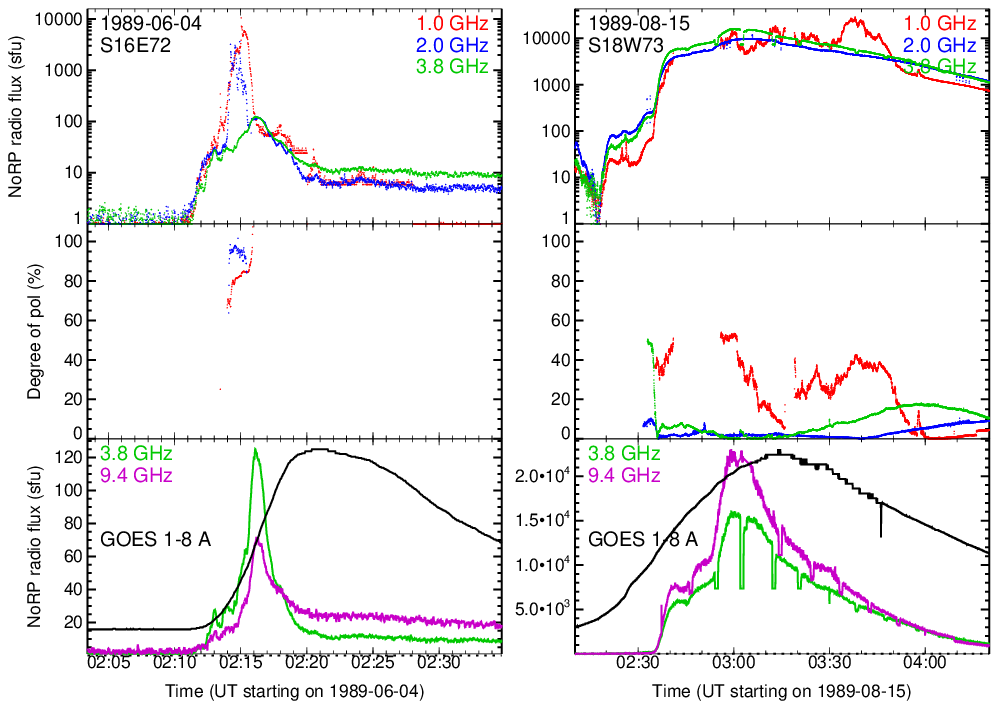}
\caption{Temporal behavior of the flares listed in Table \ref{tab_bright}. The upper panel in each case shows the 1.0 (red), 2.0 (blue) and 3.75 GHz (green) data on a logarithmic plot; the middle panel shows the corresponding (absolute) degree of circular polarization of those three frequencies at flux levels above 200 sfu; and the bottom plot shows the 3.75 (green), 9.4 (purple), 17 (orange) and 35 GHz (grey) data on a linear scale, for comparison, along with the time profile of the GOES soft-X-ray emission (black line) from the associated flare. The flare date and location (to within about 10 heliographic degrees) are shown in the upper left of each set of panels. \label{fig_flares}}
\end{figure}

\begin{figure}[t]
\figurenum{\ref{fig_flares} (continued)}
\plotone{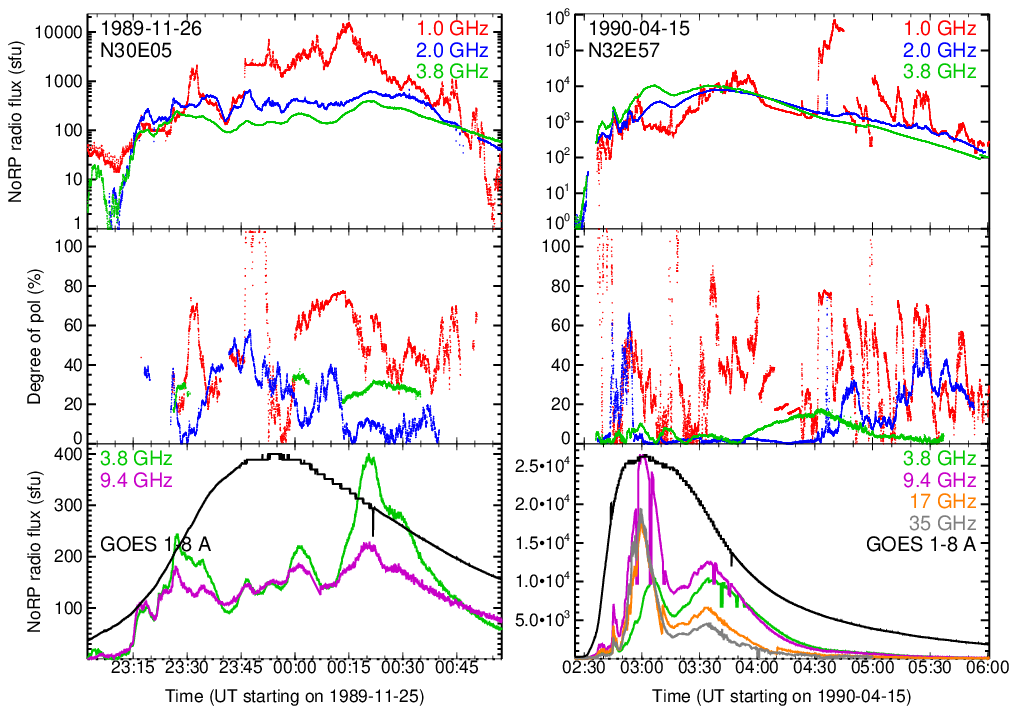}
\plotone{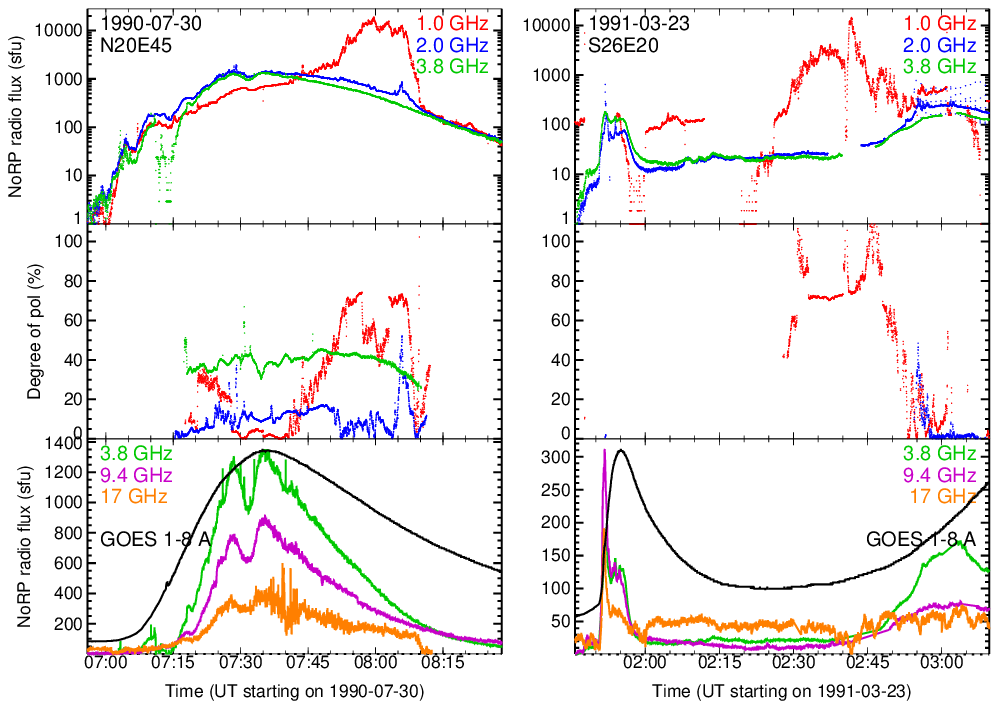}
\caption{}
\end{figure}

\begin{figure*}[t]
\figurenum{\ref{fig_flares} (continued)}
\plotone{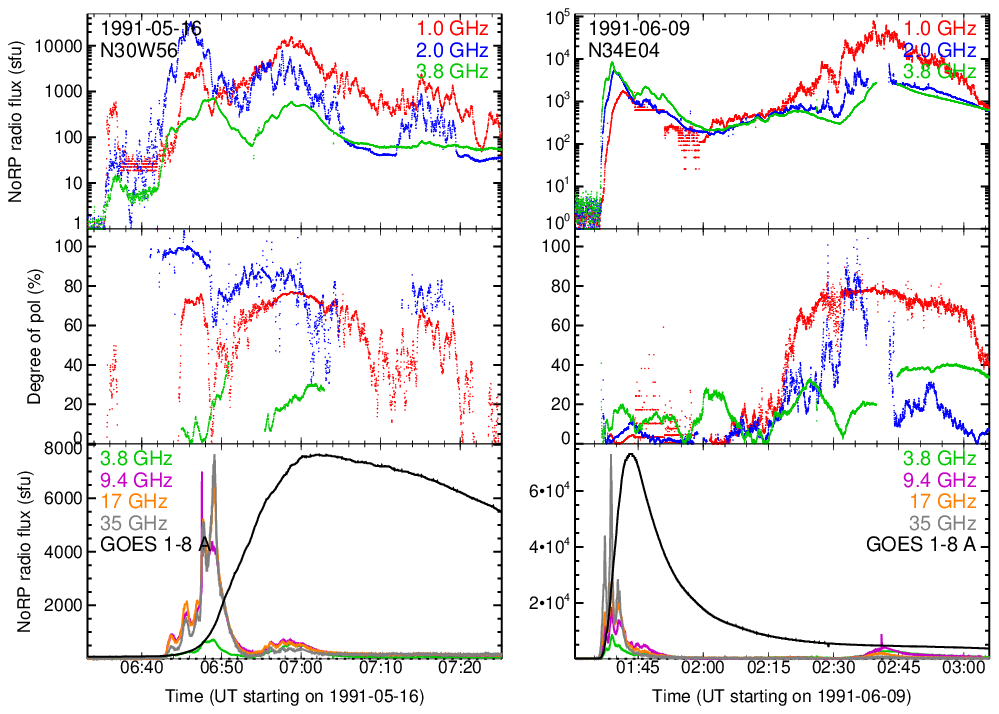}
\plotone{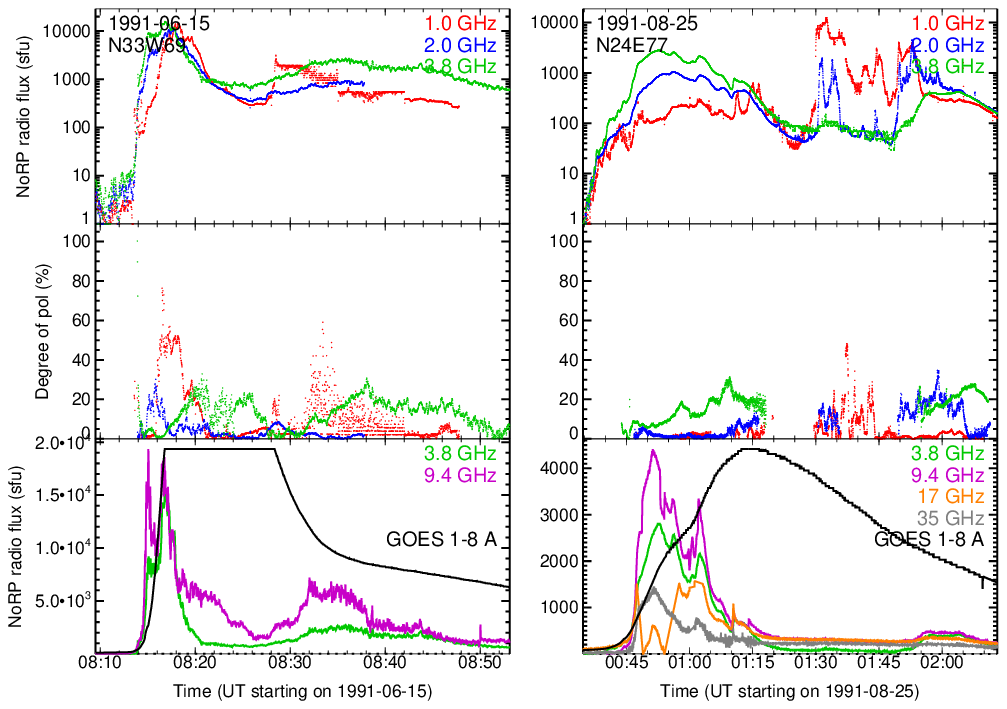}
\caption{}
\end{figure*}

\begin{figure}[t]
\figurenum{\ref{fig_flares} (continued)}
\plotone{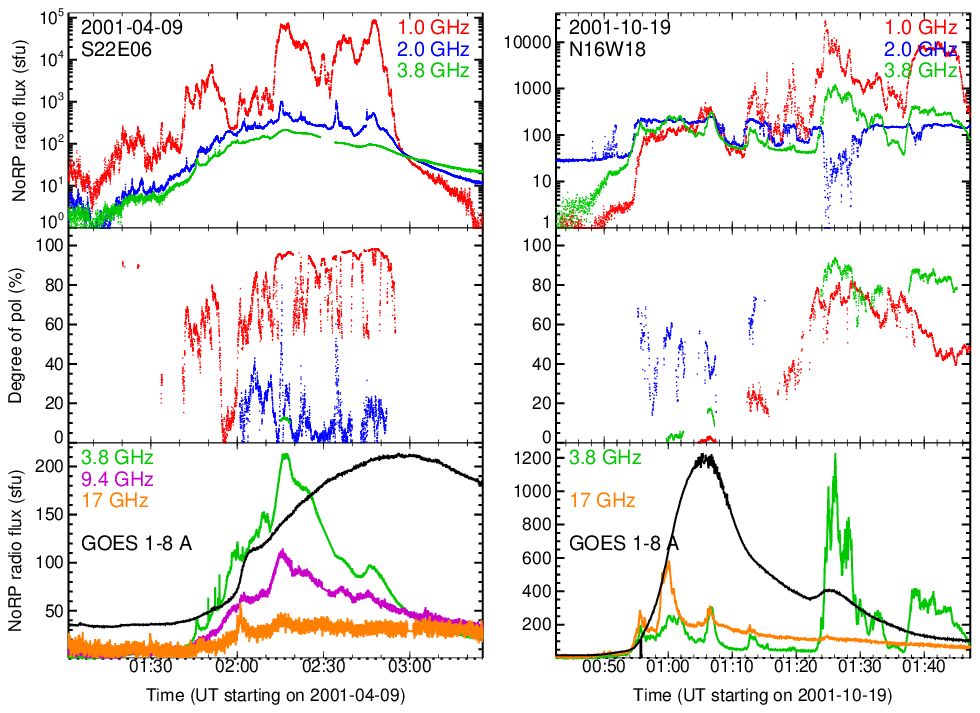}
\plotone{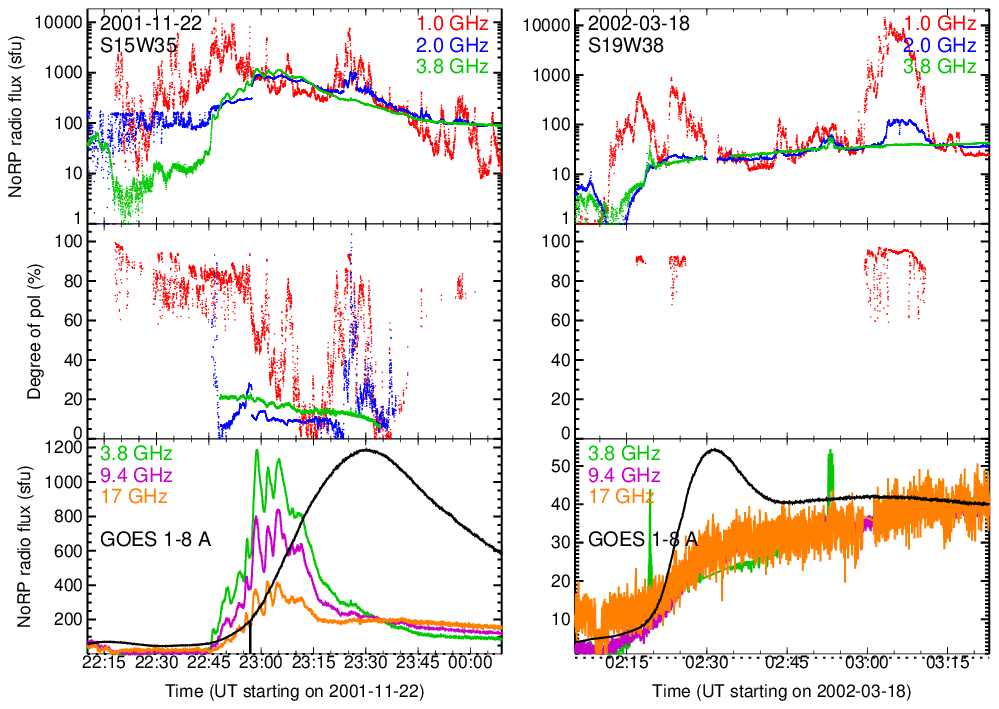}
\caption{}
\end{figure}

\begin{figure}[t]
\figurenum{\ref{fig_flares} (continued)}
\plotone{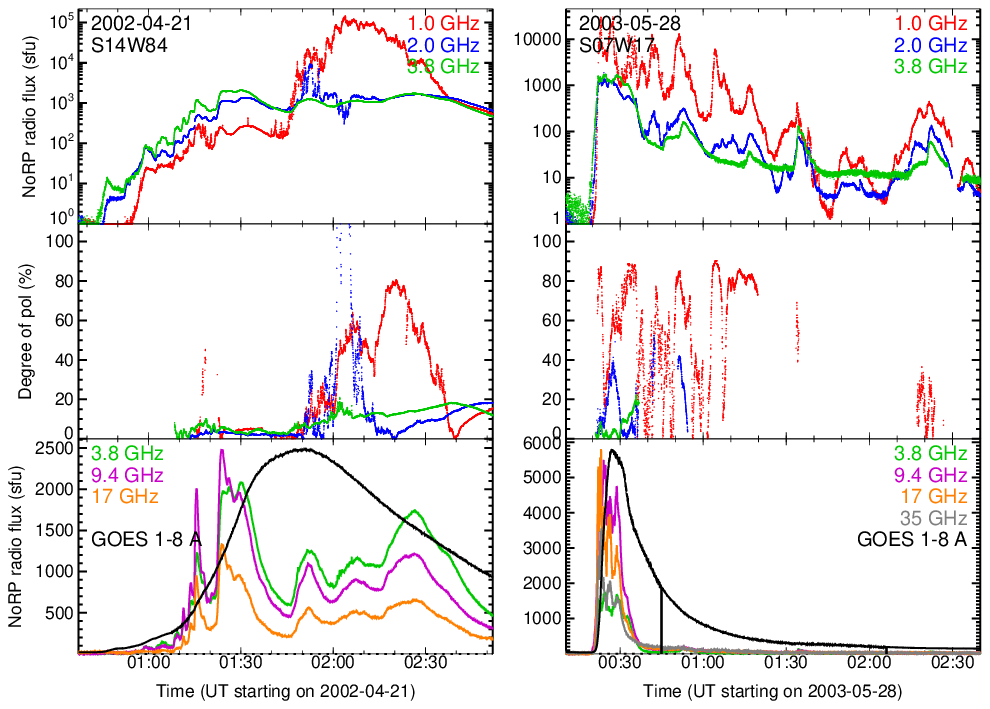}
\plotone{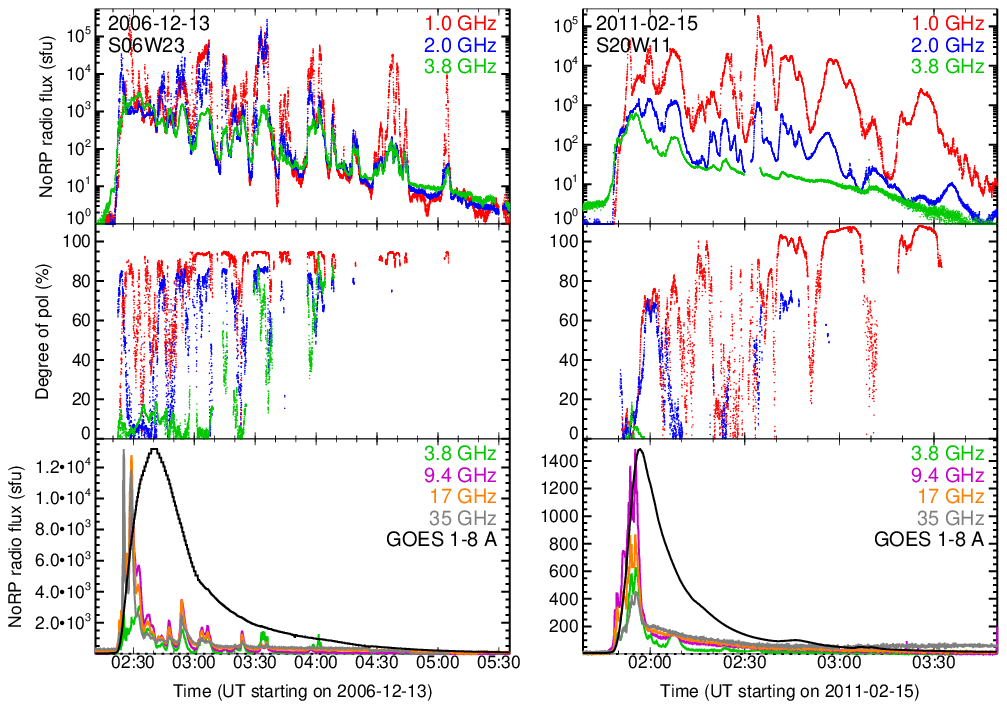}
\caption{}
\end{figure}

\begin{figure}[t]
\figurenum{\ref{fig_flares} (continued)}
\plotone{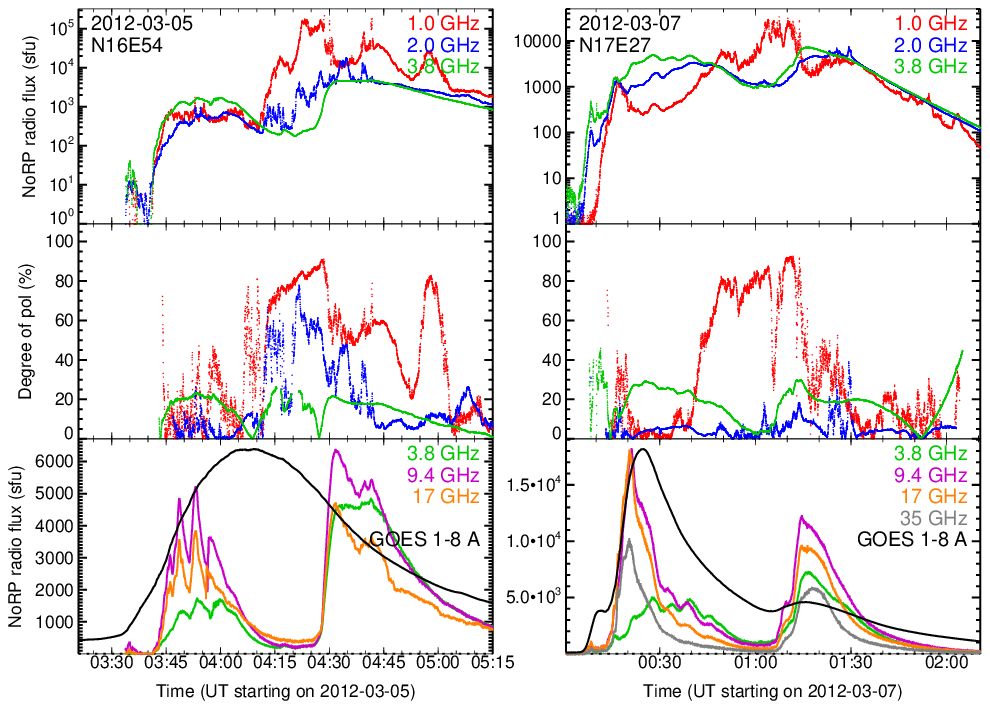}
\plotone{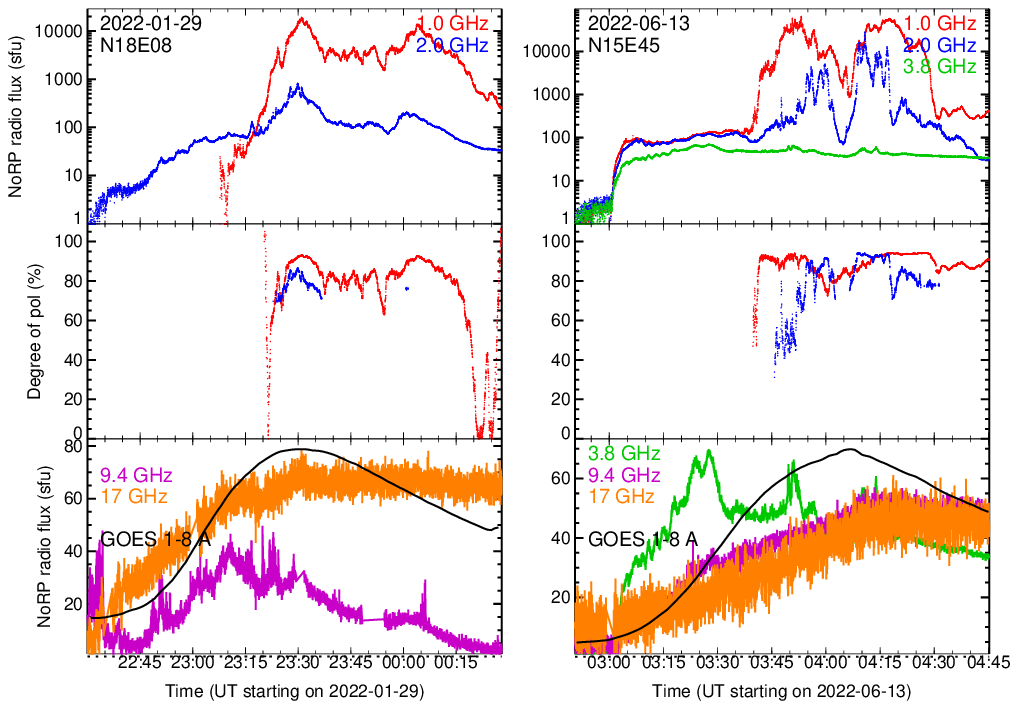}
\caption{}
\end{figure}

\end{document}